\documentclass[twocolumn]{aastex63}
\usepackage{booktabs}
\usepackage{graphicx}
\usepackage{xcolor}

\emergencystretch=\maxdimen
\shorttitle{A Sea of Black Holes}
\shortauthors{Ruiz-Rocha et al.}

\begin{document}

\title{A Sea of Black Holes: Characterizing the LISA Signature for Stellar-Origin Black Hole Binaries}

\affiliation{Department of Physics and Astronomy, Vanderbilt University, Nashville, TN 37235, USA}
\affiliation{Department of Life and Physical Sciences, Fisk University}
\affiliation{Institut für Theoretische Astrophysik, ZAH, Universität Heidelberg, Albert-Ueberle-Straße 2, D-69120,Heidelberg, Germany}
\affiliation{Physics and Astronomy Department Galileo Galilei, University of Padova, Vicolo dell’Osservatorio 3, I–35122, Padova, Italy}
\affiliation{INFN - Padova, Via Marzolo 8, I–35131 Padova, Italy}
\affiliation{TAPIR, Mailcode 350-17, California Institute of Technology, 1200 E California Blvd, Pasadena, CA 91125, USA}

\author{Krystal Ruiz-Rocha}
\affiliation{Department of Physics and Astronomy, Vanderbilt University, Nashville, TN 37235, USA}

\author{Kelly Holley-Bockelmann}
\affiliation{Department of Physics and Astronomy, Vanderbilt University, Nashville, TN 37235, USA}
\affiliation{Department of Life and Physical Sciences, Fisk University}

\author{Karan Jani}
\affiliation{Department of Physics and Astronomy, Vanderbilt University, Nashville, TN 37235, USA}

\author{Michela Mapelli}
\affiliation{Institut für Theoretische Astrophysik, ZAH, Universität Heidelberg, Albert-Ueberle-Straße 2, D-69120,Heidelberg, Germany}
\affiliation{Physics and Astronomy Department Galileo Galilei, University of Padova, Vicolo dell’Osservatorio 3, I–35122, Padova, Italy}
\affiliation{INFN - Padova, Via Marzolo 8, I–35131 Padova, Italy}

\author{Samuel Dunham}
\affiliation{TAPIR, Mailcode 350-17, California Institute of Technology, 1200 E California Blvd, Pasadena, CA 91125, USA}

\author{William Gabella}
\affiliation{Department of Physics and Astronomy, Vanderbilt University, Nashville, TN 37235, USA}

\begin{abstract}
Observations by the LIGO, Virgo and KAGRA (LVK) detectors have provided new insights in the demographics of stellar-origin black hole binaries (sBHB). A few years before gravitational-wave signals from sBHB mergers are recorded in the LVK detectors, their early coalescence will leave a unique signature in the ESA/NASA mission Laser Interferometer Space Antenna (LISA). Multiband observations of sBHB sources between LISA and LVK detectors opens an unprecedented opportunity to investigate the astrophysical environment and multi-messenger early-alerts. In this study, we report the sBHB sources that are expected to be present in the LISA data derived directly from the hydrodynamic cosmological simulation Illustris. By surveying snapshots across cosmological volume, metallicity and look-back time, we calculate the expected sBHBs present in the LISA data for various combinations of mission lifetime and stellar population model.
For stellar population estimates consistent with the LVK rates, we find that only 10 sBHBs across Illustris snapshots would be detected with significant confidence for a 10-year LISA mission, while a 4-year LISA mission would detect only $\sim 1$ sBHBs.
Our work paves the way for creating LISA mock data and bench marking LISA detection pipelines directly using cosmological simulations.
\end{abstract}


\section{Introduction} \label{section:Introduction}
The Laser Interferometer Gravitational-Wave Observatory (LIGO) made the first direct detection of gravitational waves in 2015, revealing a merger of black holes with masses 36 and 29 $M_{\odot}$ \citep{First_GW_Detection}. These black holes, dubbed {\it stellar origin binary black holes} (sBHB) by the gravitational wave community, are much more massive than the roughly 10 $M_{\odot}$ black holes in accreting binaries that had been detected electromagnetically \citep{2019_Electromagnetic_observations}.
This discovery expanded our understanding of black hole populations, providing insight into their formation and evolution \citep{2016Abbot_astrophysicalimplications, 2016Abbot_gw150914}. However, the formation channel of these binaries is still under debate. The current favored mechanisms involve formation via: a) isolated binaries in the galactic field \citep{1998BetheBrown, 2002Belczynski,2010Banerjee,2014MennekensVanbeveren,2016deMinkMandel,2016Marchant,2018Giacobbo,2018Kruckow,Shao_2021,Broekgaarden2022} ; b) dynamical interactions within dense stellar environments  \citep{2000PortegiesZwart,2010Banerjee,2014Ziosi,2015Rodriguez,2016Rodriguez,2016Mapelli,2017Banerjee,2017Askar,2021Rodriguez}; and c) binaries within gaseous AGN accretion disks \citep{2018McKernan, 2017Stone,2017Bartos, 2022Ford}.

For the Laser Interferometer Space Antenna (LISA), an ESA/NASA space-based gravitational wave observatory that was adopted in 2024 and is set to launch in 2035\citep{2017_LISA, redbook}, sBHBs are an important class of sources. LISA is expected to be sensitive from around $10^{-4}$ to 0.1~Hz, a frequency range that captures the inspiral phase of the sBHB population. A vast majority of sBHBs would be detected as Galactic binaries that are millions of years away from merger~\citep{Belczynski_2010,Nelemans01,Liu_2014,Breivik_2020,Sesana2020, 2022Wagg, 2023_Babak}. Here, we discuss a subclass of sBHBs that are multi-band sources; these are a few years or less from merger, and could be observed subsequently or even jointly with ground-based gravitational-wave detectors ~\citep{2016_Multi-BandSesana, 2019BAAS...51c.109C, 2019_Gerosa, 2020_Karan}.

Multiband observations alone enable a more accurate measurement of masses, spins, eccentricity, and distance~\citep{Vitale_2016, Randall_2021, Toubiana_2021, 2024arXiv240611926R}, but their unique power is in providing better spatial resolution and enough lead time before merger to conduct rigorous observational campaigns to identify electromagnetic counterparts \citep{LISA_White_Paper_2020, 2019_Lamberts, Korol_2017, Digman_2023}. Multiband data is vital to differentiate between sBHB formation channels as well as to constrain their final evolutionary sequence \citep{2016_Breivik, LISA_White_Paper_2020, LISA_Proposal_2017, 2016Nishizawa_a, 2017Nishizawa_b}.

\begin{figure}[t!]
    \centering
    \includegraphics[width=0.5\textwidth]{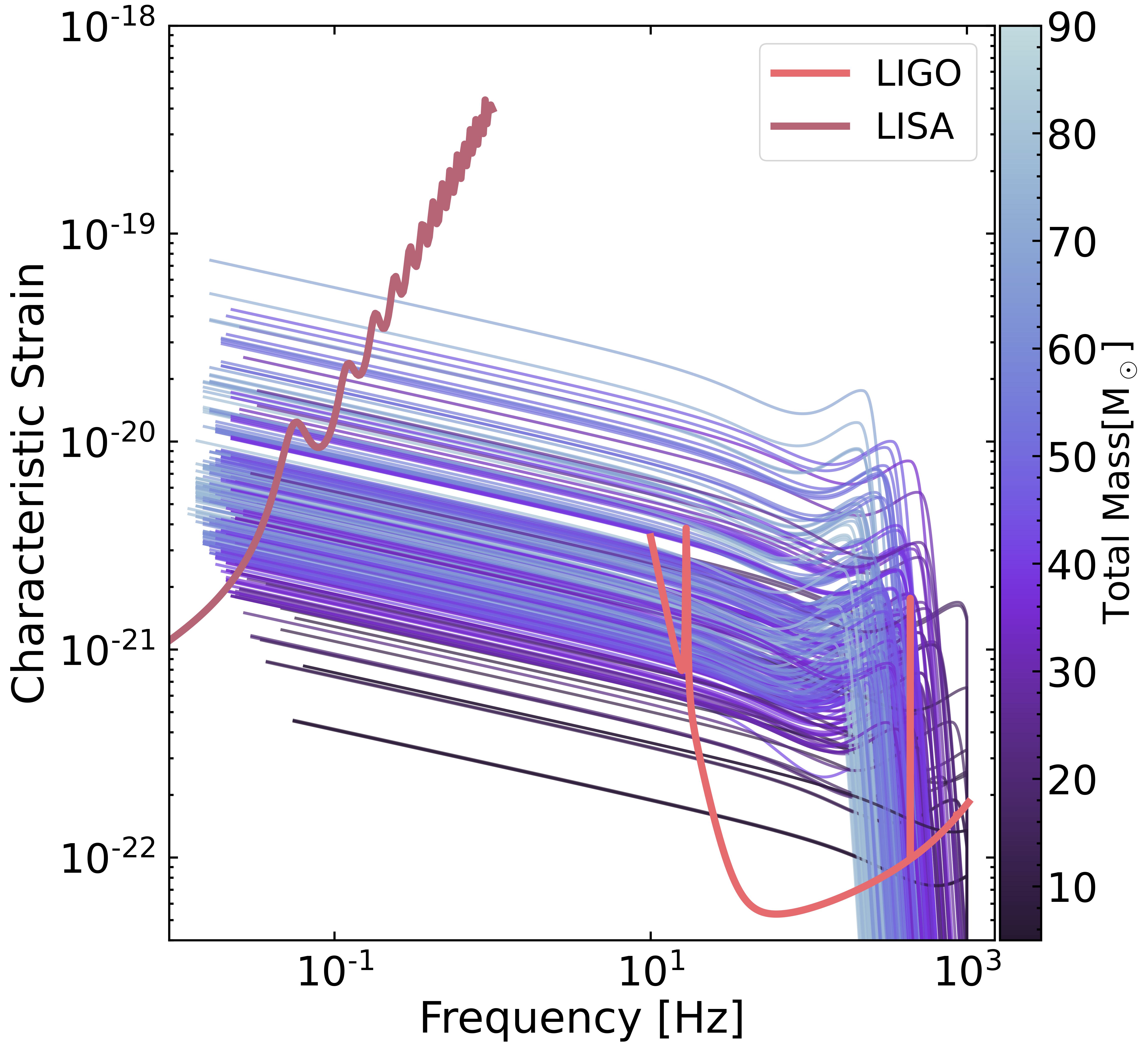}
    \caption{Characteristic strain vs. frequency for sBHBs formed within the snapshot at redshift z = 13.4 ~\citep{Mapelli}. We only show a subset of these sBHBs that eventually merged under $z\leq0.5$. The strain from each sBHB is colored by the total binary mass in the source frame. The left curve line is the adopted LISA sensitivity curve~\citep{2019Robson}, the right curve is LIGO's sensitivity curve\citep{2020LIGOCurve}.}
    \label{fig:Characteristic Strain}
\end{figure}

To investigate the possibility that stellar origin binaries arise in an isolated binary stellar population, \citet{Mapelli} used the cosmological hydrodynamic simulation suite Illustris \citep{2014_Vogelsberger_b} to predict the astrophysical and LIGO-observable rate of sBHB mergers, given a range of binary stellar population models. Illustris provided a self-consistent time-evolving map of star-formation rate, mass, and metallicity to be used as input to a binary population synthesis code, making it possible to paint a stellar population onto the simulation, let it evolve, and track the formation and ultimate merger of sBHBs.

Our work extends this study, identifying those sBHB mergers within the Illustris simulation volume that would be observable with LISA. Formally, the time to merge for sBHBs in the LISA band can be from months to millions of years, but we narrowed our investigation to those sBHBs that transition from LISA to ground-based detectors within the lifetime of the LISA mission; these sources are expected to be loudest in the LISA band and are therefore the most promising for multiband and multi-messenger observations \citep{2019_Gerosa}.
Figure \ref{fig:Characteristic Strain} shows the characteristic strain for a subset of sBHB mergers calculated with pycbc \citep{pycbcv240} within the \cite{Mapelli} database that may be seen by LISA as shown by the left noise curve from \cite{2019Robson} and merge within LIGO as shown on the rightmost noise curve from \cite{2020LIGOCurve}. These binaries are chosen from the snapshot at the highest redshift(z = ~ 13.4), because it contains sBHBs that merge within a wide variety of lookback times, showing the robustness of \cite{Mapelli}'s database for studying multi-messenger astronomy.

The paper proceeds as follows: Section \ref{section:Methods} describes our methods, including a brief review of the simulation, the stellar population input from ~\citet{Mapelli}. In addition, section \ref{section:SNR} discusses the signal-to-noise calculations for our binary database. In section \ref{section:Results} we discuss the results, including the multiband rates (section \ref{subsection:MultibandRate}), and number of detectable multiband sources (section \ref{subsection:MultibandExpectedNumber}). Finally, the conclusion and discussion are within section \ref{section:Conclusion} where we summarize our results, point out caveats, compare to other results, and identify areas for future work.

\section{Methods} \label{section:Methods}

Illustris is a set of cosmological hydrodynamic N-body simulations run using the AREPO moving mesh code \citep{2014_Vogelsberger_b, 2010_AREPO}. To model galaxy formation and evolution in a cosmological context, the simulation self-consistently treats dark matter, supermassive black holes, gas, and stars, including prescriptions for gas cooling, star formation, metal enrichment, supernovae and associated feedback, as well as massive black hole formation and evolution (including gas accretion, mergers, and several modes of feedback). Illustris is a cosmological volume of $106.5^3$ Mpc$^3$ initialized in a Wilkinson Microwave Anisotropy Probe (WMAP-9) cosmology with the following parameters \citep{2013_wmap9}: $\Omega_M = 0.2726$, $\Omega_{\Lambda} = 0.7274$, $\Omega_b =0.0456$, h = $0.704$,  and $H_0 = 70.4$ kms$^{-1}$Mpc$^{-1}$, where $\Omega$ represents the mean density relative to the critical density of the Universe, and $M$, ${\Lambda}$, and $b$ refer to matter, the cosmological constant, and baryons, respectively. $h$ is the dimensionless Hubble parameter \citep{2013_Croton}, and $H_0$ the Hubble constant. The Illustris suite is composed of 6 simulations: {\sc Illustris-(1,2,3)}, which are the hydrodynamical simulations modeling galaxy formation and include all the star formation, gas and massive black hole physics described above, and {\sc Illustris-DM-(1,2,3)}, which are the dark matter-only variant. We use Illustris-1 for its higher mass resolution; the Illustris-1 simulation contains $1.6 \times 10^{12}$ particles over all snapshots, with a dark matter mass resolution of $6.3 \times 10^6 M_{\odot}$, a baryonic matter mass resolution of $1.3 \times 10^6 M_{\odot}$ and a softening length of $\sim 710$ pc for stars and supermassive black holes (SMBHs)~\citep{2014_Vogelsberger_b,2014_Vogelsberger_a}.

Although the Illustris simulation suite is state-of-the-art, it cannot resolve individual stars and stellar remnants. Therefore, \cite{Mapelli} maps a population of sBHB, each stellar particle in snapshots spanning redshifts 0--16 using an updated version of the public binary stellar population synthesis code, Binary Star Evolution (BSE) \citep{2000_Hurley,2002_Hurley} called {\texttt MOBSE} \citep{2018Giacobbo}. For more detail about the procedure to seed Illustris with sBHB, please see \citet{Mapelli}; we briefly describe the process here. \cite{Mapelli} constructed a library of 72 distinct binary stellar population models, each 'book' in the library consisting of millions of sBHBs representing a particular metallicity and a set of assumptions about the less well-constrained binary-evolution physics, such as mass loss in the common-envelope phase, treatment for Hertzsprung gap donors, models for natal kicks, and SNe model.

\begin{figure}[t!]
    \centering
\includegraphics[width=0.44\textwidth]{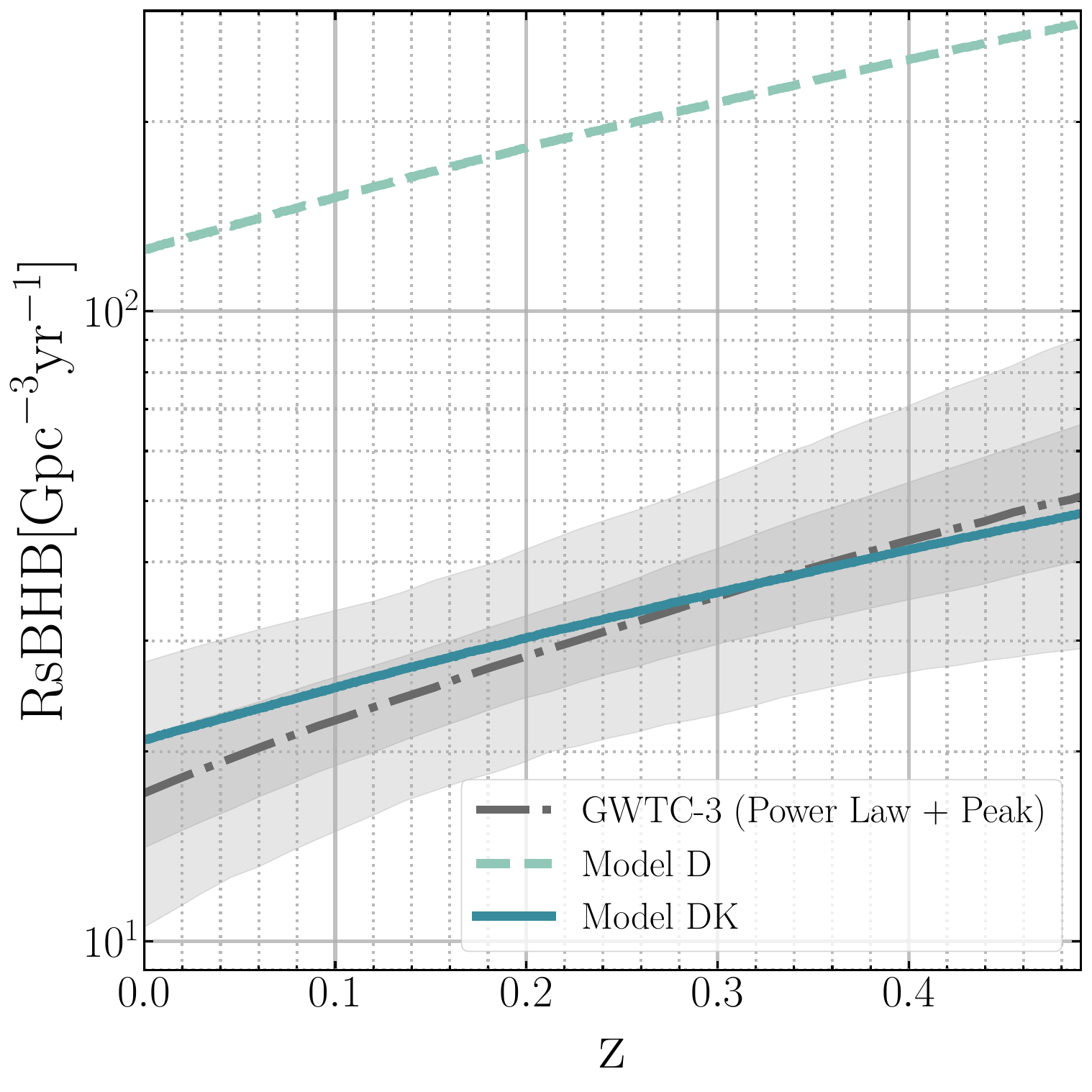}
    \caption{Rate versus redshift for the fiducial model D and DK. The dotted grey line and contours are a reproduction of LIGO'S O3 rate from \cite{2021_LIGO_O3}}
    \label{fig:Rate Comparison}
\end{figure}
 
Using this library, a sBHB distribution is assigned to Illustris-1 star particles. The metallicity of a star particle determines which stellar population 'book' to use, and the mass of the star particle determines how many sBHB to select. Each sBHB is characterized by its masses, separation, and eccentricity, which is used to calculate the binary merger timescale. By following only the sBHB that merge within a Hubble time, this earlier work resulted in a simulated census of the astrophysical merger rates for these sources, as well as estimates of the number of binaries and mass distribution of sBHB mergers that LIGO will see for various stellar population models, as shown in Figure 6 of \cite{Mapelli}. 

Note that for our paper, we adopt models D and DK as our fiducials; these assume a delayed supernova model, a treatment of Hertzsprung gap (HG) donors, and natal kicks for compact binaries \citep{Mapelli}. Model D was the fiducial model for the \citet{Mapelli} study, and it matched the LIGO rates O1. However, more recent LIGO observations indicate a lower sBHB merger rate that is better represented by model DK, so we included that model as well. The difference between model D and DK is that D employs a method described in \cite{2012_Fryer} to rescale natal kicks for black holes to take into account the fraction of the stellar envelope that can fall back onto the compact object during the explosion, $f_b$, which can range from 0 to 1 \citep{2012_Fryer,2015_Spera}, as shown by equation \ref{kicks}
\begin{equation}\label{kicks}
    v_{BH} = v_{NS}(1-f_b),
\end{equation}

\noindent where $v_{NS}$ is the velocity distribution in \citep{Hobbs_2005}. Model DK applies the velocity distribution found by  \cite{Hobbs_2005} for natal kicks of neutron stars to the black holes.

\begin{figure}[t!]
    \centering
   \includegraphics[width=0.45\textwidth]{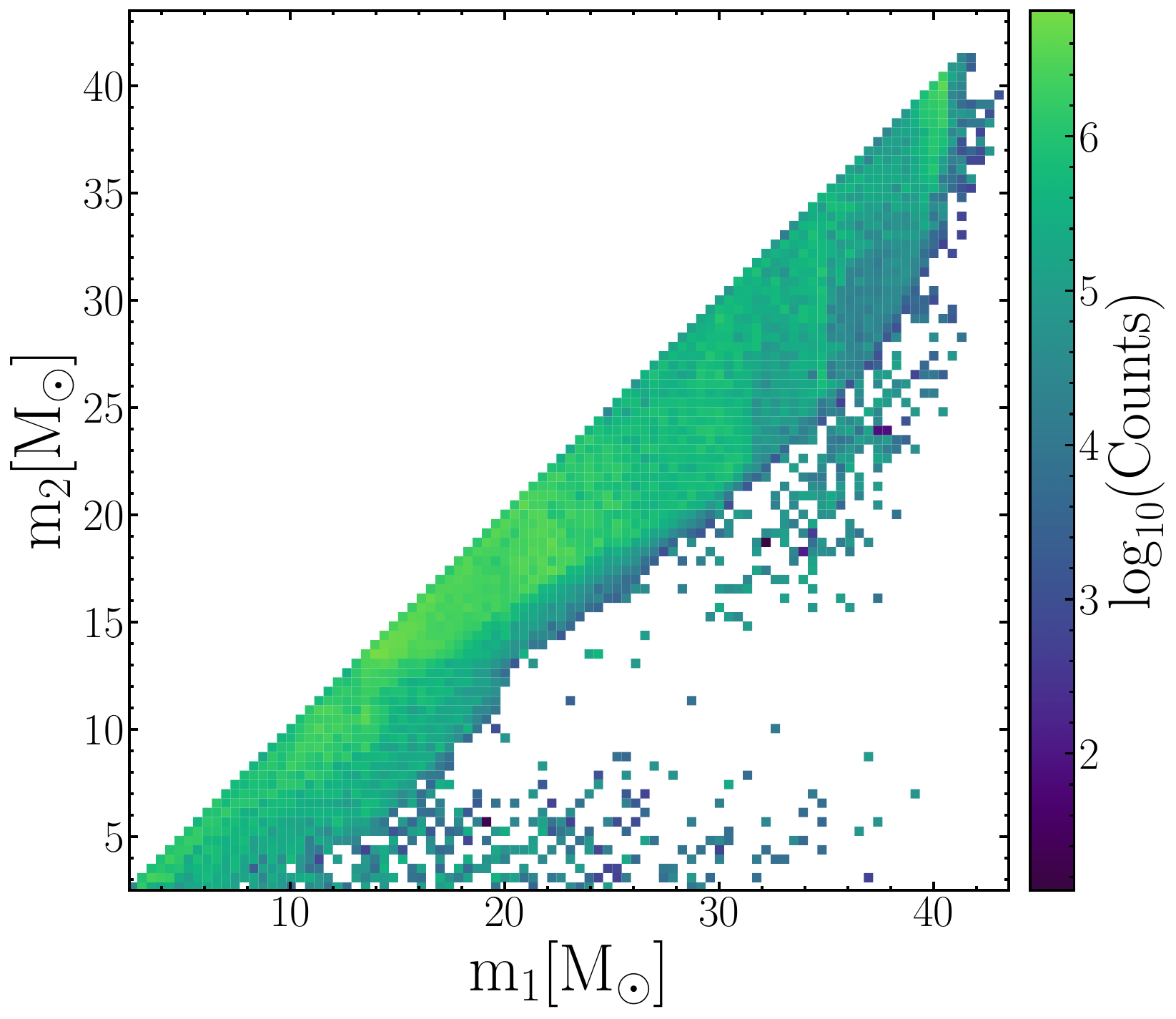}
    \caption{Heatmap showing the mass distribution of sBHBs for the first black hole mass ($m_1$) and second black hole mass ($m_2$) of Model D of the Mapelli post-processing files, restricted to sBHBs merging at redshifts below 0.5. In the original database, $m_1$ was not always greater than $m_2$; however, in this figure, $m_1$ has been set to be greater than $m_2$ for all sBHBs. The colors represent the logarithm of the sBHB counts in each bin. For Model DK, the mass heatmap exhibits a similar distribution to Model D  for $m_1$ and $m_2$, but the sBHB counts are approximately 6 times lower across all bins.}
    \label{fig:Heatmap}
  
\end{figure}

The sBHB merger rate in model D is $181$ $\mathrm{Gpc}^{-3}\mathrm{yr}^{-1}$ at a redshift of 0.2. At the time of publication, this rate was the most consistent with O1 LIGO observations. However, LIGO data from its observing run 3 (O3) places the sBHB merger rate between 17.9 and 44 $\mathrm{Gpc}^{-3}\mathrm{yr}^{-1}$ at a redshift of 0.2 \citep{2021_LIGO_O3}. This filters into our results as an overestimate of the LISA and LIGO multiband rate when we use model D. In contrast, model DK offers a much closer match to O3 rates with a sBHB merger rate of $29$ $\mathrm{Gpc}^{-3}\mathrm{yr}^{-1}$ at redshift 0.2, approximately 6 times smaller than the rate of model D. This similarity is shown in Figure \ref{fig:Rate Comparison} which plots the merger rates for model D, DK, and LIGO'S rate for O3 against redshift. The dashed line represents model D's rate and the lower line is model DK's rate. LIGO's rate is shown by the dotted gray line with the inner contours containing 50\% of the data and the outer contours containing 90\% of the data. Model DK's rate consistently falls within the 50\% contours across the full redshift range and aligns more closely to LIGO's rate for $z > 0.3$. We note that the star formation rate history in the \textsc{Illustris-1} simulation is  overall in good agreement with observations \citep{2017Madau}, except for an overestimate of $\sim{40\%}$ at low redshift ($z \lesssim{0.2}$). However, this overestimate has a negligible impact on the merger rate density of sBHBs \citep{Mapelli}, because only stars with nearly solar metallicity form at such a low redshift: the efficiency of BHB mergers drops by about three orders of magnitude if the metallicity $Z$ is $>0.2$~Z$_\odot$ \citep{2018Giacobbo}.

\subsection{Calculating the Signal-to-Noise Ratio in the LISA Band} \label{section:SNR}

We began with 108 snapshots ranging from redshift $0.0072$ to $13.6$, each consisting of a database of the sBHB formed at that redshift. This database contains information on: 1. the progenitor stellar particle ID from Illustis; 2. metallicity; 3. black hole component masses; 4. formation redshift; 5. sBHB merger time.

Since each snapshot contains only those sBHB that formed within a particular snapshot time interval, the varying sBHB merger time implies that a sBHB at a previous snapshot may still exist at a given redshift, but will be missing in the snapshot; indeed, the sBHBs seeded in a given snapshot will merge in the future over a wide range of timescales. It is worth reiterating that the database only contains sBHBs that merge within a Hubble time. Consequently, sBHBs in the simulated population that do not merge by the present day are missing from the current database. We leave the analysis of the quasi-static sBHB population for a future paper.The quasi-static population would serve to increase the population statistics and our predictions, however these binaries may not have high enough SNR to be observed beyond the Galaxy or local volume.

\begin{figure}[t!]
    \centering
    \includegraphics[width=0.5\textwidth]{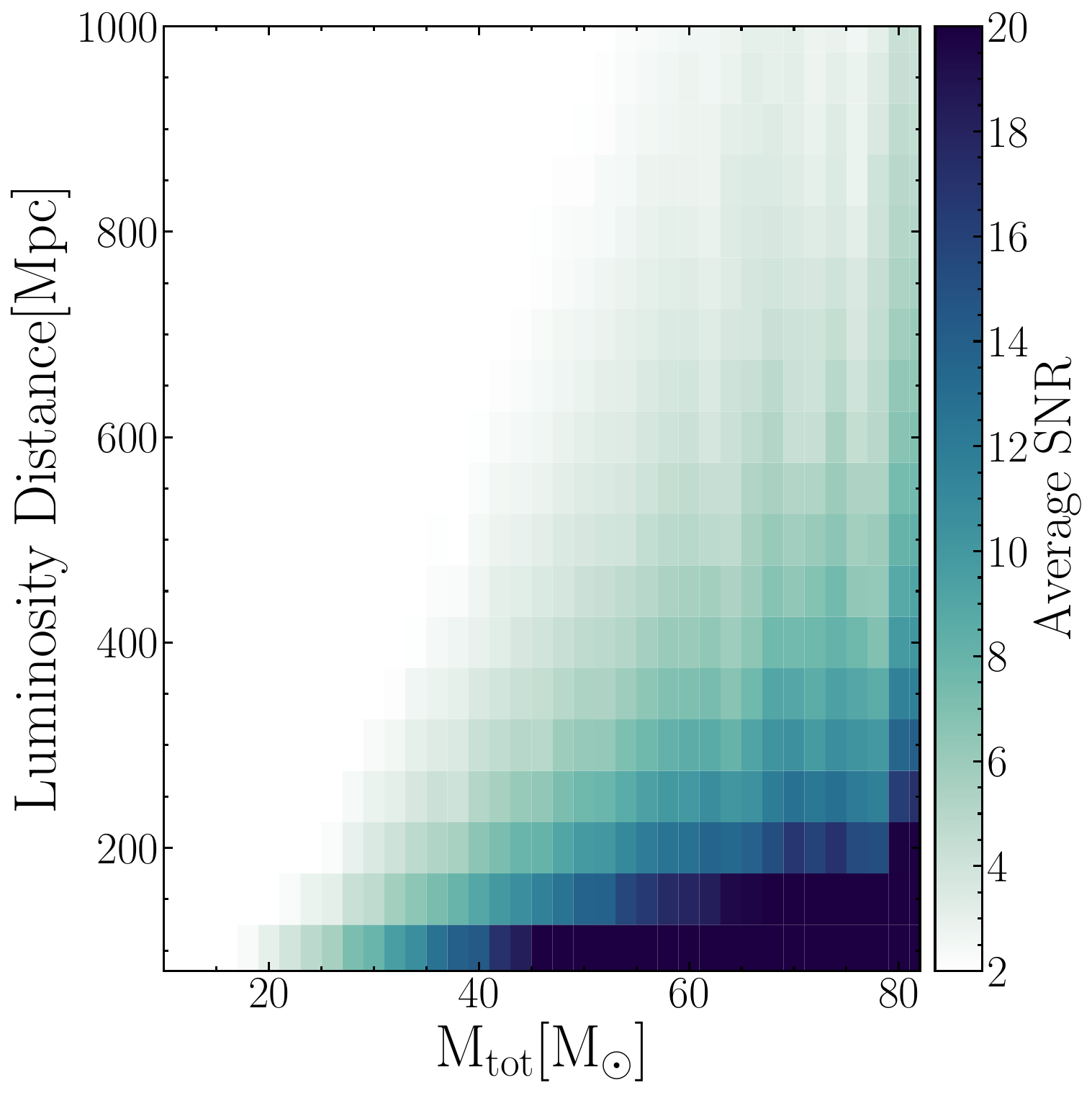}
    \caption{Heatmap of the total mass and luminosity distance colored by the average SNR in each grid for an expected LISA mission lifetime of 10 years.}
    \label{fig:SNR Heatmap}
\end{figure}

\begin{figure*}[ht!]
    \centering
    \includegraphics[width=0.7\textwidth]{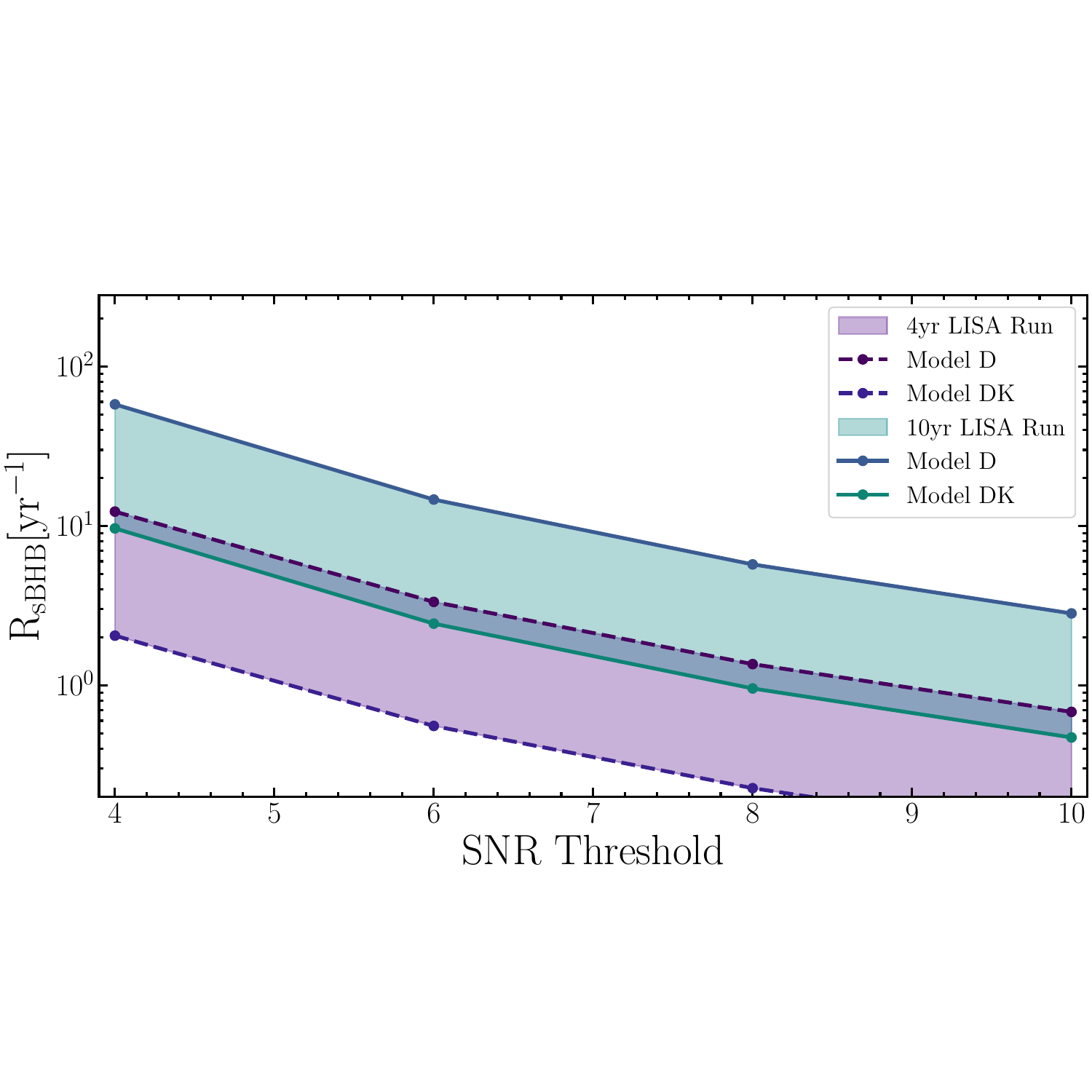}
    \caption{Multiband rates versus SNR threshold for our sBHBs database by integrating over redshift for equation \ref{eq:expectedrate} for both Model D and Model DK for a 4-year LISA mission, shown with purple contours, and a 10-year LISA mission shown with the teal contours.}
    \label{fig:rates}

\end{figure*}

To identify sBHB mergers that will be observable by both LISA and LIGO, we first restrict the database to those sBHBs that merge within a horizon redshift of 0.5; this decision greatly reduced the number of sBHBs in the Illustris volume from $7.72 \times 10^9$ to $1.19 \times 10^9$. Our choice is conservative and encompasses the expected horizon distance for sBHBs (see Figure 2 of \citet{2020_Karan}). Figure \ref{fig:Heatmap} displays the black hole mass distribution for those systems that merge within a Hubble time and merge below a redshift of 0.5 in model D. For this figure, we also forced $m_1$ to always be greater than $m_2$, which was not the case in our database. In these models, the most common sBHB merger arises between black holes with commensurate masses of roughly 15--25 $M_\odot$. The heatmap for Model DK is similar to Figure \ref{fig:Heatmap}, except the counts are smaller by a factor of $\sim 6$.
We caution that neither model D nor DK contain sBHBs with component masses $m_{\rm BH} > 43 M_\odot$. LIGO has certainly observed black holes more massive than the upper limit of these models \citep{2019Abbot,2021Abbot,2023Abbot,2024Abbot}. It may well be that the high mass end of the observed BHB merger mass spectrum represents additional processes, such as hierarchical merging, that may be at play and are therefore not well-represented by BSE \citep{2000_Hurley,2002_Hurley,2018Giacobbo}. We leave a multiband treatment of the higher mass black holes observed by LIGO to a future paper.

Using the redshift-limited database, we extract the masses and lookback time for each black hole binary merger and determine the gravitational wave frequency, $f_{\rm obs}$, of these binaries at a time $T_{\rm obs}$ before the merger. This calculation assumes quasi-circular binaries and can be described by the following equation:

\begin{equation}
    f_{\rm GW} =  \frac{c^3}{8\pi G M_c}\left( \frac{5GM_c}{c^3} \right)^{3/8}\frac{1}{T_{\rm obs}^{3/8}},
    \label{eq:beginning frequency}
\end{equation}

\noindent where c is the speed of light, G is the gravitational constant, $M_c$ is the chirp mass, and $T_{\rm obs}$ is the time until the BHB merges in seconds \citep{2018Chan,Bassan_2014}. We adopt 4 and 10 years for $T_{\rm obs}$, the projected duration of science operations for the LISA mission.  As outlined in \cite{2019_Gerosa}, this choice produces a louder SNR in the LISA band while enabling a relatively contemporaneous multiband source.
(\cite{2019_Gerosa} considered a $T_{\rm wait}$, the time between detecting an event with a space-based and a ground-based detector. In our work, $T_{\rm obs}$ is the same as $T_{\rm wait}$.)
The masses in equation \ref{eq:beginning frequency} are in the detector frame, and are converted from the source frame found in the database to detector frame masses using: $m_{\rm detector} = m_{\rm source}(1+z)$, where z is the sBHB merger redshift.

\section{Results} \label{section:Results}

\subsection{Multiband Rate for detectable sBHBs}  \label{subsection:MultibandRate}

With this information we can then calculate the expected signal-to-noise ratio (SNR). For this, we use the python {\texttt LISA Sensitivity} code from \citet{2019Robson} and also available on GitHub at: \href{https://github.com/eXtremeGravityInstitute/LISA_Sensitivity}{LISA Sensitivity}, for a given an observation duration and cosmology. This code models the adopted LISA mission sensitivity to laser shot noise, acceleration noise, and test-mass force noise over a mission duration, and calculates the characteristic strain, and SNR of a binary given the component masses, distance, and an optional sky location and inclination. Note that unless specified, the SNR is averaged over inclination angle, polarization, and sky position. Additionally, since the sBHBs did not include spin, the gravitational wave signal is approximated by a phenomenological waveform model for non-spinning binaries, referred to as PhenomA \citep{2007Ajith}. In the database, the LISA SNR for a observation run of 10 years technically ranged from 0 to $1\times 10^7$, with the anomalously high SNR populated by a sBHB at a luminosity distance of less than a kpc. Figure~\ref{fig:SNR Heatmap} maps the average SNR as a function of total mass and luminosity distance assuming a mission duration of 10 years. In this figure we see, as expected, that nearby high mass binaries yield the highest SNR. For example sBHB system with $M_{\rm tot} = 80 \, M_\odot$ will visible up to 200 Mpc in LISA with SNR $\gtrsim20$. Such high-mass binaries will be visible in LISA out to ${\sim}1$~Gpc with an average SNR ${\sim}6$. The luminosity distance for high-mass sBHB in Figure~\ref{fig:SNR Heatmap} at a SNR threshold of 8 is consistent with the multiband detection radius reported in~\cite{2020_Karan}.

    \begin{figure*}[t!]
    \centering
    \includegraphics[width=0.85\textwidth]{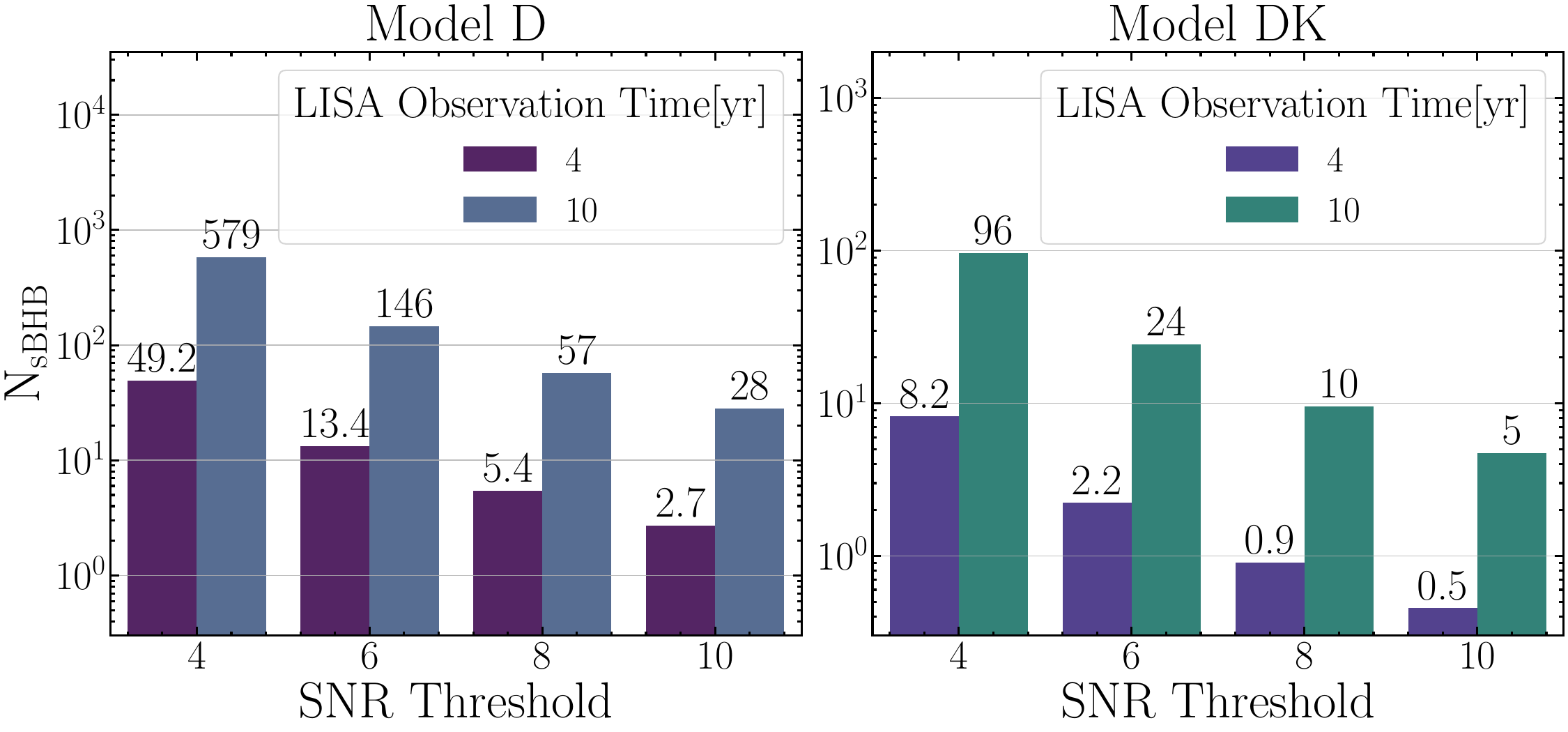}
  
    \caption{The expected number of multiband sBHBs calculated for different SNR thresholds. The left panel assumes model D and a 4 or 10-year LISA mission, while the right panel model assumes model DK, which better approximates the O3 LIGO event rate.}
    \label{fig:MultibandNumber}
    \end{figure*}

Now that we have the SNR for each sBHB in hand, we can determine the multiband event rates and the expected number of detections for LIGO and LISA as a function of SNR. We calculate the expected astrophysical rate of a population above a certain SNR threshold using equation \ref{eq:expectedrate} from \citep{2022Chen}:
\begin{equation}\label{eq:expectedrate}
    \frac{dN}{dzdt}=\frac{d^2 n(z,\rho)}{dzdV_c}~\frac{dz}{dt}~\frac{dV_c}{dz}~\frac{1}{1+z}
\end{equation}

\noindent where $V_c$ is the co-moving volume, $n(z)$ is the number of sBHBs at a redshift z, $\frac{1}{1+z}$ converts $\frac{dz}{dt}$ to the observer frame, and $\frac{d^2 n(z,\rho)}{dzdV_c}$ is the rate found in the simulation volume above a certain SNR threshold. This can be found directly from the simulation, expressed by:

\begin{equation}\label{eq:simrate}
    \frac{d^2 n(z,\rho)}{dzdV_c} = \frac{N(z,\rho)}{\Delta z V_{sim}},
\end{equation}

\noindent where $N(z,\rho)$ is the number of sBHBs merging within a redshift bin, $\Delta z$ is the width of the redshift bin and $V_{\rm sim}$ is the simulation volume, which in our case is $(106.5 {\rm Mpc})^3$.

The multiband rate of sBHB mergers, equation \ref{eq:expectedrate}, is then integrated over for each SNR threshold over redshift. The rate for each SNR threshold is shown in Figure \ref{fig:rates} for 4-year (purple) and 10-year (teal) LISA mission for model D and model DK. During a 4-year observational run, our results for model D indicate that LISA and LIGO will detect $12$ sBHBs ${\rm yr^{-1}}$ with SNR $\geq 4$. However, this number drops to $0.68$ ${\rm yr^{-1}}$ when we increase the SNR threshold to 10 for model D. Model DK also shows the same pattern but is six times smaller than model D, i.e $2$ ${\rm yr^{-1}}$ for SNR $\geq 4$ to 0.11 ${\rm yr^{-1}}$ for an SNR $\geq 10$. If we extend the observational run to 10 years, the detectable multiband rate increases to $58$ ${\rm yr^{-1}}$ with an SNR threshold of 4 to $3$ ${\rm yr^{-1}}$ for SNR $ \geq 10$ for model D. For model DK we see only $10$ ${\rm yr^{-1}}$ for SNR $ \geq 4$ to $0.5$ ${\rm yr^{-1}}$ for SNR $ \geq 10$.

\subsection{Expected Number of Multiband sBHBs} \label{subsection:MultibandExpectedNumber}

Converting the integrated multiband rate from equation \ref{eq:expectedrate} to the number of detectable sBHBs, we used:
\begin{equation}\label{eq:number}
    N = \frac{dN}{dt} T_{\rm obs},
\end{equation}
\noindent where $\frac{dN}{dt}$ is the integrated rate across redshift for a SNR threshold calculated from equation \ref{eq:expectedrate} and $T_{\rm obs}$ is the LISA mission lifetime.
Figure \ref{fig:MultibandNumber} displays the number of sBHB that could be detected by both LIGO and LISA given a 4- and 10-year LISA mission lifetime. We see once again that the detectable multiband BH rate decreases as the SNR thresholds increase for both a 4-year and 10-year mission. For a 4-year LISA run, model D (left panel) exhibits a detectable number of multiband sBHB mergers that ranges from 2.7--49.2 as the SNR threshold decreases from 10 to 4, while model DK (right panel) yields between 0.5--8.2. For a 10-year mission, the number of sBHB mergers in model D ranges from 28--579, and from 5--96 for model DK.

From our results, we conclude that the multiband detection rates (per year) for a 10-year LISA mission increases by ${\sim}4$x for both the models across all SNR thresholds when compared to a 4-year LISA mission. This is consistent with an average increase in SNR of sBHB by around $\sqrt{10/4}$ between the two LISA mission lifetimes, thus increasing the detection volume by a factor of ${\sim}4$. Similarly, the total number of multiband detections in a 10-year LISA mission is anticipated to be ${\sim}10$x higher at SNR threshold of 10 and ${\sim}12$x higher at SNR threshold of 4 when compared to 4-year LISA mission. The mild increase at low SNR thresholds compared to high SNR thresholds is due to an increase in the intrinsic sBHB merger rates with redshift in the local universe~\citep{Mapelli}.  
\begin{table*}[htbp]
    \centering
    \caption{
    Expected number of multiband sBHBs for a range of LISA mission durations. The terms marginal and expected are defined by SNR $\geq 4$ and SNR $\geq8$, respectively. For each range, the lower value represents model DK, while the higher value represents model D.}
    \begin{tabular}{@{}ccc@{}}
        \toprule
        \multicolumn{1}{c}{\begin{tabular}[c]{@{}c@{}}\textbf{LISA mission}\\ \textbf{lifetime}\end{tabular}} & \multicolumn{1}{c}{\begin{tabular}[c]{@{}c@{}}\textbf{Marginal}\\ \textbf{sBHB detections}\end{tabular}} & \multicolumn{1}{c}{\begin{tabular}[c]{@{}c@{}}\textbf{Confirmed}\\ \textbf{sBHB detections}\end{tabular}} \\
        \midrule
        \ \ \ \ 4 years & $8.2 - 49.2$& $0.9 - 5.4$\\
        \ \ \ \ 6 years & $24.2 - 145$ & $2.6 - 15.3$ \\
        \ \ \ \ 8 years & $52.5 - 315.2$ & $5.4 - 32.5$ \\
        \ \ \ \ 10 years & $96.5 - 578.9$& $9.6-57.4$\\
        \bottomrule
    \end{tabular}

    \label{tab:lisa_detections}
\end{table*}
\subsection{Comparison to Previous Work}
\label{subsection:PreviousWork}

Previous work has been done to estimate sBHB multiband rates and expected numbers for LISA~\citep{2019_Gerosa, 2016_Multi-BandSesana, 2017_Multi-BandSesana,2016_Multi-BandKyutoku, 2023_Multi-BandZhao}.
\cite{2016_Multi-BandSesana,2017_Multi-BandSesana} presented a semi-analytic data-driven approach using LIGO observations from O1, assuming two underlying black hole mass distributions and intrinsic merger rates: a Salpeter mass function with a merger rate of $100 \,{\rm Gpc^{-3}yr^{-1}}$, and log-uniform mass function with a rate of $35 \, {\rm Gpc^{-3}yr^{-1}}$. These higher merger rates inferred from O1 naturally lead to a larger number of predicted multiband events compared to our models.

\cite{2016_Multi-BandKyutoku} examined the multiband detection of GW150914-analogs assuming an O1 event rate as well, however this work was primarily focused comparing the effect of different noise curves for eLISA, a precursor to the mission architecture adopted by LISA. For a 10 year mission, they predicted that the number eLISA-LIGO multiband sBHB detections ranged from 8--400, depending on the mission configuration.

\cite{2019_Gerosa} updated the data-driven approach of the Sesana work using LIGO/Virgo/KAGRA observations from O1 and O2; the Salpeter mass function yielded a merger rate of $57^{+40}_{-25} \,{\rm Gpc^{-3}yr^{-1}}$, while the log-uniform yielded a rate of $19^{+13}_{-8.2} \, {\rm Gpc^{-3}yr^{-1}}$. For a 10-year LISA mission, their finding of 4-22 multiband LISA+LVK detections above an SNR of 8 is most consistent with our model DK, while model D is roughly 6 times higher.
Additionally, \cite{2019_Gerosa} created binary population synthesis simulations with spinning black hole binaries for isolated systems that evolved through the common-envelope phase. They perform 7 simulations, each varying the natal kicks assigned at BH formation. Their natal kicks were distributed using a Maxwellian distribution with a dispersion between 0 and 265 km/s, and the expected multiband detection rates were found to potentially vary up to a factor of $\sim 30$ depending on the natal kick choice.  Model DK is most comparable to their model with a dispersion of $265$ km/s and is slightly larger than their predictions.

In a related analysis, \cite{2023_Multi-BandZhao} created a Monte Carlo-sampled database of mock sBHB based on GWTC-3 observations with a rate of $19^{+8.4}_{-8.5}$ $\mathrm{Gpc}^{-3}\mathrm{yr}^{-1}$. For a 4-year LISA mission and an SNR threshold of 8 they predict 3-15 events, which falls within the range of both of our models.

\section{Conclusion} \label{section:Conclusion}

In this paper, we predicted the expected rate and signature of sBHB that will be detectable by both LISA and LVK, with the hope that the early warning of a merger will help constrain sky location and help facilitate an alert of an impending merger for electromagnetic and ground-based GW detectors. Combining two binary population synthesis models with a cosmological hydrodynamic simulation, we found the multiband rate and expected number of sBHB detectable as a function of SNR and LISA mission duration. Table~\ref{tab:lisa_detections} summarizes our findings for four different mission durations. In this table, we define marginal detections as those with SNR $\geq 4$, while confirmed sBHB detections must have SNR $\geq8$. Model DK produces the minimum value in each column while model D yields the maximum value. We note that model DK, the binary population synthesis model with larger supernova natal kicks, yields overall sBHB merger rates that are more consistent with LVK observations through O3. Although it may be tempting to conclude that small black hole natal kicks can be ruled out by LVK observations, the number of competing open questions in binary stellar evolution -- from the strength of core overshooting, to the impact of unstable mass transfer, to the core-collapse explosion process itself -- preclude the ability to do so. Indeed, many binary stellar population models exist, with entirely different choices in the late stage binary evolution physics~\citep[e.g.][]{Broekgaarden22}. Advances in binary stellar population modeling will allow a better mapping of the unknown parameter space in binary stellar evolution and will therefore better enable comparisons to gravitational wave observations~\citep{SEVN}.

Our work provides a novel approach for building LISA mock data catalogs of sBHB directly using cosmological simulations. Such synthetic catalogs provide useful benchmark for LISA data-analysis pipelines~\citep{Babak_2010, 2022arXiv220412142B, Lackeos_2023}. To maximize the LISA SNR of a given sBHB, we deliberately modeled those that would merge over a timescale that is roughly the LISA mission duration; there are certainly sBHBs that will merge in a ground-based gravitational wave band hundreds of years in the future, but these sources are much fainter in the LISA band when so far from merger, and are therefore not ideal multiband sources.
We also neglected sBHBs that did not merge by the end of the simulation; this population of monochromatic, or nearly monochromatic, sBHBs may well be buried in the confusion foreground of Galactic compact object binaries~\citep{redbook,Wagg22}

This study considers only the field sBHB formation channel, yet LVK observations may hint that multiple formation channels could be at play~\citep{Bourranais21}. Cosmological simulations of the scale needed to make universal predictions are as yet unable to resolve the competing mechanisms mentioned in the introduction: AGN disks and dense stellar systems such as nuclear star clusters and globular clusters. We will explore building a framework to seed a simulation with sBHB from multiple formation channels in a future work.\

We also note that the ability to make population inference using sBHB may be hampered by parameter estimation; there is a degeneracy between eccentricity and spin precession, for example, that may make parameter estimation of multiband sBHB particularly tricky\citep{Romero23}. This underlines the importance of continued developments in analysis techniques specifically for the LISA sources that are less subject to such degeneracies, but are currently missing. Future work will investigate the influence of higher masses, spin magnitudes, spin-orbit precession and orbital  eccentricity of sBHBs on LISA detection rates.

\section*{Acknowledgments}
KHB acknowledges and appreciates support from NSF NRT-2125764. M.M. acknowledges financial support from the European Research Council for the ERC Consolidator grant DEMOBLACK, under contract No. 770017 (PI: Mapelli) and from the German Excellence Strategy via the Heidelberg Cluster of Excellence (EXC 2181 - 390900948) STRUCTURES. This work used the resources provided by the Vanderbilt Advanced Computing Center for Research and Education (ACCRE), a collaboratory operated by and for Vanderbilt faculty at Vanderbilt University.

\bibliography{sample63}{}
\bibliographystyle{aasjournal}

\end{document}